\documentclass[12pt,preprint]{aastex}
\slugcomment{18 SEP 2012}

\shorttitle{CO Mapping of IRAS~22272+5435}
\shortauthors{Nakashima et al.}

\begin{document}
%%\begin{CJK*}{UTF8}{gbsn}

%%%%%%%%%%%
%% title %%
%%%%%%%%%%%

\title{CO Structure of the 21 Micron Source IRAS 22272+5435:\\ A Sign of a Jet Launch?}

%%%%%%%%%%%%%%%%%%%%%%%%%
%% authors information %%
%%%%%%%%%%%%%%%%%%%%%%%%%

\author{Jun-ichi Nakashima\altaffilmark{1}, Nico Koning\altaffilmark{2}, Nikolaus H. Volgenau\altaffilmark{3}, Sun Kwok\altaffilmark{1}\\ Bosco H. K. Yung\altaffilmark{1}, Yong Zhang\altaffilmark{1}}

\altaffiltext{1}{Department of Physics, University of Hong Kong, Pokfulam Road, Hong Kong, China\\ Email(JN): junichi@hku.hk}

\altaffiltext{2}{Department of Physics and Astronomy, University of Calgary, Calgary, Canada T2N 1N4}

\altaffiltext{3}{California Institute of Technology, Owens Valley Radio Observatory, Big Pine, CA 93513, USA}

%%\end{CJK*}

%%%%%%%%%%%%%%
%% abstract %%
%%%%%%%%%%%%%%

\begin{abstract}
We report the results of radio interferometric observations of the 21-$\mu$m source IRAS~22272+5435 in the CO $J=2$--1 line. 21-$\mu$m sources are carbon-rich objects in the post-AGB phase of evolution which show an unidentified emission feature at 21~$\mu$m. Since 21-$\mu$m sources usually also have circumstellar molecular envelopes, the mapping of CO emission from the envelope will be useful in tracing the nebular structure.  From observations made with the Combined Array for Research in Millimeter-wave Astronomy (CARMA), we find that a torus and spherical wind model can explain only part of the CO structure. An additional axisymmetric region created by the interaction between an invisible jet and ambient material is suggested.
\end{abstract}

%%%%%%%%%%%%%%
%% keywords %%
%%%%%%%%%%%%%%


%%%%%%%%%%%%%%%%%%
%% Introduction %%
%%%%%%%%%%%%%%%%%%

\section{Introduction}
A number of carbon-rich proto-planetary nebulae (PPNe) -- objects in the evolutionary transition phase between the asymptotic giant branch (AGB) stars and planetary nebulae (PNe) -- are found to exhibit a strong unidentified emission feature at 21 $\mu$m \citep{kwo89}.  Although over twenty years have passed since the initial discovery of this feature, the chemical origin of the carrier has yet to be identified.  The nebular morphological structures of known 21-$\mu$m sources have been studied  in the optical with the Keck telescope and Hubble Space Telescope \citep[{\it HST}; see, e.g.,][]{uet00,uet01}, and in the infrared by the Very Large Telescope \citep{lag11}. 

Interferometric observations of molecular rotational lines (particularly the CO lines) are useful for investigating the morphological properties of circumstellar envelopes of 21~$\mu$m sources. However, the number of objects for which structure can be resolved by conventional radio interferometers is limited. As of 2012, IRAS 07134+1005 is the only object that has been well-resolved by interferometric observations in the CO lines \citep{mei04,nak09}. The CO observations show that a torus was likely formed by an equatorially enhanced mass-loss event in the last 2500--3000 years, but there is no evidence of a jet \citep{nak09}. Since, in many PPNe/PNe, bipolar jets exhibit a shorter dynamical timescale than tori \citep{hug07}, the structure of IRAS 07134+1005 suggests that 21~$\mu$m sources are transient objects between the torus and jet formation phases.  Investigations of the morpho-kinematic properties of other 21~$\mu$m sources would help to test this hypothesis.

In this paper, we report the results of CO observations of IRAS 22272+5435 in the CO $J=2$--$1$ line, using the Combined Array for Research in Millimeter-wave Astronomy (CARMA). Since the object is located at a high declination and the angular size of the infrared torus is relatively small (roughly 3$''$), the observation is possible uniquely only with CARMA. In Section 2, we briefly summarize previous morpho-kinematic studies of IRAS 22272+5435. We give details of the present observations and data reduction in Section 3. In Section 4, we summarize the observational results. In Section 5, we analyze the data using the morphokinematic modeling tool {\it Shape}, and we discuss the consequences of the models constructed by {\it Shape} in Section 6. Finally, we summarize our main results in Section 7.

%%%%%%%%%%%%%%%%%%%%%%%%%%%%%%%%%
%% Summary of previous observations of IRAS 22272+5435%%
%%%%%%%%%%%%%%%%%%%%%%%%%%%%%%%

\section{Summary of Previous Observations of IRAS 22272+5435}

IRAS 22272+5435 (= HD 235858 = SAO 34504) was first proposed as a PPN candidate soon after its detection by the Infrared Astronomical Satellite (IRAS), based on its relatively strong fluxes in both the optical and infrared bands \citep{pot88,hri91}. Subsequent optical spectroscopic observations classified the central star as spectral type G5 Ia \citep{hri91}. Its carbon-rich nature is based on the detections of C$_2$ and CN molecular bands in the atmosphere of the central star \citep{hri91,hri95}. The 21~$\mu$m feature was discovered in the IRAS low resolution spectrum \citep[LRS;][]{kwo89}. The distance to IRAS 22272+5435 is estimated to be 1.67~kpc, based on a dust radiative transfer model fit \citep{szc97}. We adopt this value in our analysis.

The morphology of IRAS 22272+5435 has been studied at various wavelengths. Mid-IR images at arcsecond and subarcsecond resolutions show an elongated emission core \citep{mei97,day98,uet01}. The elongation is interpreted as the result of an inclined dust torus or disk. High-resolution optical images obtained by $HST$ reveal a reflection nebulosity of very faint surface brightness with a clear view of the star at the center of the nebula \citep{uet00}. The optical nebulosity is elongated approximately perpendicular to the core elongation as seen at the mid-IR images. Near-IR polarimetry by \citet{gle01} separates the polarized (i.e. dust scattered) emission from the unpolarized (i.e. direct) stellar emission. Their $J$-band polarized image shows a ring-like structure embedded in an elongated halo. \citet{uet01} suggest, on the basis of their dust radiative transfer modeling, that the central star left the AGB about 380 yr ago, after the termination of the superwind, and has been experiencing post-AGB mass-loss,  with a sudden, increased mass ejection about 10 yr ago.

CO emission from IRAS 22272+5435 was first detected in the $J=1$--0 line at the Five College Radio Astronomy Observatory \citep{zuc86} and in the $J=2$--1 line at the James-Clerk-Maxwell Telescope \citep[JCMT; ][]{woo90}.  Further observations have been made in both lines by \citet{ner98} using the IRAM 30m telescope. \citet{hri00} and \citet{hri05} observed the CO $J=2$--1, 3--2 and 4--3 lines using the JCMT and Heinrich Hertz Submillimeter Telescope (HHT). The spectra are well fitted by a single parabolic profile in the $J=2$--1 line and by a Gaussian profile in the CO $J=4$--3 line; there is no indication of high-velocity wings in either spectrum. \citet{buj01} also found no high-velocity wings in the CO $J=1$--0 and 2--1 lines. \citet{hri05} fit a one-dimensional radiative transfer model to their single-dish spectra, and found that  the CO $J=2$--1 line is matched best by a $r^{-4}$ density law. A $r^{-3}$ dependence is acceptable, but a $r^{-2}$ dependence is clearly too flat-topped when compared to the observed spectrum. The CO $J=1$--0 line is also best matched by a $r^{-4}$ density model, but the predicted intensity is about 30\% less than the observed peak. A $r^{-2}$ density model clearly does not fit the observed line shape, suggesting that the envelope is disturbed by post-AGB asymmetric mass-loss. 

Interferometric observations in the CO $J=1$--0 line were first made with BIMA \citep{kwo97, fon06}. The morphology of the molecular gas revealed by the BIMA observation in the CO $J=1$--0 line are roughly consistent with a spherically expanding envelope with an angular size of about 20$''$, even though on smaller scales (2--3$''$) the envelope seems to slightly deviate from spherical symmetry.

%%%%%%%%%%%%%%%%%
%% Observation and data reduction%%
%%%%%%%%%%%%%%%%%

\section{Details of Observation and Data Reduction}
The CARMA observations of IRAS 22272+5435 were made in the D configuration on March 12, 2009 and in the C configuration on April 18, April 19, May 29, and November 7, 2009. Most observations were made under good atmospheric conditions, with the exception of the data from April 18 and May 29, which required extensive flagging, but still contained good data. The total on-source integration time was 12.0 and 3.9 hours in the C and D configurations, respectively. CARMA comprises 15 telescopes (6 $\times$ 10.4~m, 9 $\times$ 6.1~m), with baselines ranging from 30 to 350~m in C configuration and from 11 to 150~m in D configuration. The half-power beam-widths (HPBWs) are 50$''$ for the 6~m antennas and 30$''$ for the 10~m antennas at the frequency of the CO $J=2$--1 line. The phase center of the map was R.A.$=22^{\rm h}29^{\rm m}10.37^{\rm s}$, decl.$=+54^{\circ}51'06.4''$ (J2000).

CARMA's 3-band spectral correlator was configured with two 500 MHz bands (with a spectral resolution of 31.25 MHz) and one 31 MHz band (with a spectral resolution of 0.49 MHz). The CO $J=2$--1 line ($\nu_{rest} = 230.538000$ GHz) was placed in the center of the upper sideband (USB) of the 31 MHz band, yielding a velocity resolution of 0.64~km~s$^{-1}$. The velocity coverage across the 31 MHz band is about $40$~km~s$^{-1}$. The 500 MHz bands were set to frequencies {\it away} from the CO line, in order to measure the continuum emission. Observations of IRAS 22272+5435 were interleaved about every 20 minutes with a nearby gain calibrator, BL Lac, to track the phase variations over time. 

The data were calibrated using the MIRIAD software package \citep{sau95}. Absolute flux calibration was determined from observations of Mars, Neptune and MWC349, and we estimate a flux accuracy of $<$~30\%. The level of uncertainty has two causes: (1) different primary calibrators were used for the different observing trials, and (2) the flux of the gain calibrator, BL Lac, was independently verified to fluctuate over the time span that our observations were made. Image processing of the data was also performed with MIRIAD. All calibrated visibility data were combined using the MIRIAD task {\it uvaver} prior to transforming the data into the image plane. The continuum emission was removed from the line emission map by fitting a baseline to the line-free channels and then subtracting the baseline with MIRIAD's {\it uvlin} task. The robust weighting scheme (we applied ''robust=0.5'') yielded a clean beam of 1.1$''$$\times$1.0$''$ and a position angle of $-80.0^{\circ}$. The continuum emission was mapped integrating over a roughly 1~GHz range (2 $\times$ 500 MHz correlator windows: the exact frequency ranges for the continuum observation were 224.82209 GHz -- 225.29084 GHz and 225.32217 GHz -- 225.79092 GHz), and the continuum flux was measured by fitting a two dimensional Gaussian function (using the MIRIAD task {\it imfit}). The measured total integrated flux of continuum emission is 1.1~Jy. The continuum emission source was not spatially resolved with our synthesized beam.

%%%%%%%%%%%%%%%%%
%% Results %%
%%%%%%%%%%%%%%%%%

\section{Results}
In Figure~1, we present the total intensity profile of the $^{12}$CO $J=2$--1 line. The integral area for creating the spectrum is a $9'' \times 13''$ box centering around the phase center. As previously reported \citep{buj01,hri05}, the line profile exhibits a parabolic shape with no high-velocity components. The peak intensity is 21.4~Jy at $V_{\rm LSR}=-29.9$~km~s$^{-1}$. The systemic velocity obtained by fitting a parabolic function is $V_{\rm LSR}=-28.1$~km~s$^{-1}$ (with a peak of 21.4~Jy). The line-width at the zero intensity level is 21.0~km~s$^{-1}$ (corresponding to an expanding velocity of 10.5~km~s$^{-1}$). The integrated intensity is 279.0~Jy~km~s$^{-1}$. This value is different from the value obtained in  previous single-dish measurement of 486.5~Jy~km~s$^{-1}$ \citep[][assuming a conversion factor of HHT of 35~Jy~K$^{-1}$]{hri05}. If we assume 25--30\% of uncertainty in the HHT flux, the discrepancy in fluxes could be interpreted by the uncertainty in flux measurements. However, the discrepancy, of course, may suggest that some of the flux emitted from large-scale structure is resolved out in the CARMA observations. Anyway, the line profiles from the CARMA observations and HHT observations are almost exactly the same, suggesting that both profiles capture the source's essential kinematic properties.

Figure~2 shows the total intensity map of the continuum-subtracted CO~ $J=2$--1 line emission superimposed on a map of the 1~mm continuum emission. Because the continuum emission is not resolved, subtracting it from the CO emission does not have an effect on the morphological information of the line. In the CO image, we clearly confirm the double intensity peaks northwest and southeast of the phase center. The angular separation of the two intensity peaks is about 1.4" (corresponding to $3.5 \times 10^{16}$~cm at the distance of 1.67~kpc). The position angle of the line passing through the two intensity peaks is about 120$^{\circ}$. The central resolved structure is surrounded by a roughly spherical component, but its outer regions are elongated to the north-east and south. The 3~$\sigma$ and 7~$\sigma$ contours exhibit a deviation from spherical, while contours above a 11~$\sigma$ level exhibit a roughly spherical pattern. The morphology seen in the CO~ $J=2$--1 line is different from that of the CO~$J=1$--0 line \citep{fon06}; the BIMA observations in the CO~$J=1$--0 line did not clearly resolve the central structure, even though they also confirmed the spherical component, which is more extended than the structure seen in the $J=2$--1 line.

In Figure~3, we present the total intensity map of the CO~$J=2$--1 line superimposed on the mid-infrared (MIR) 12.5 $\mu$m image \citep[left panel;][]{uet01} and $HST$ $I$--band image \citep[right panel;][]{uet00}. The central region is enlarged. The MIR image exhibits two intensity peaks like the CO image, but interestingly, the lines connecting the MIR and CO peaks are almost perpendicular. This result contrasts with IRAS~07134+1005 \citep{nak09}, in which the CO structure is nearly coincident with the MIR structure of a rim-brightened torus. The observation of IRAS~22272+5435 in the $J=2$--1 line may trace a relatively lower temperature region than the CO $J=3$--2 observations in IRAS~07134+1005 made by \citet{nak09}. Since \citet{uet01} reasonably fit the MIR images by a rim-brightened torus, CO emission detected in the present observation seems to originate from components other than a rim-brightened torus. In the right-panel of Figure~3, the CO contours show a weak correlation with the optical protrusions suggested by \citet{uet01}. Although the direction of the central bipolar structure does not correspond to the directions of the four protrusion, outer contours surrounding the central bipolar structure seem to exhibit a correlation with the elliptical protrusions. In particular, the largest protrusion toward the south-east shows a relatively good correlation with the CO contours. \citet{uet01} pointed out that the directions of the optical protrusions are strikingly coincident with the directions in which there are fewer dust grains. Therefore, one may think that the ultraviolet radiation of the central star, which is leaked from the fewer-dust-region, could play a role to disturb the CO intensity distribution. 

Figures~4  and  5 show the channel velocity maps of the CO~$J=2$--1 line. In figure~4, we present the entire emission region, while in figure~5,  the enlarged central region is presented together with the HST $I$-band image and the locations of emission peaks suggested by \citet{uet01}. In Figure 4, we find that the size of the outer spherical component increases as the velocity comes close to the systemic velocity at $-28.1$~km~s$^{-1}$. This tendency suggests that the outer component surrounding the central structure is interpreted with a spherically expanding flow.  In Figures~4 and 5, we see many intensity peaks around the map center. Even though in total intensity maps (Figures~2 and 3) two intensity peaks stand out at the north-west and south-east of the phase center, the channel velocity maps reveal that there exist intensity peaks on the equatorial plan of a rim-brightened torus suggested in previous MIR imaging (the blue crosses in Figure~5 represent the location of two emission peaks found in a mid-infrared image, corresponding to brightened rims of a torus). For example, in channels of $-$33.7 km~s$^{-1}$,  $-$28.6 km~s$^{-1}$, $-$26.1 km~s$^{-1}$, $-$24.8 km~s$^{-1}$ and $-$22.3 km~s$^{-1}$, intensity peaks on the equatorial plane of the torus are clearly seen.  Since the locations of CO intensity peaks on the equatorial plan of the torus is relatively farther away from the map center compared to those of the mid-infrared peaks, presumably the CO~$J=2$--1 line traces the somewhat outer part (i.e., lower temperature part) of the torus compared to the mid-infrared emission. On the other hand, we also see intensity peaks on the symmetric axis of the suggested torus. Such intensity peaks on the symmetric axis are seen in almost every channel: in particular, clear features are seen in channels of $-$32.4 km~s$^{-1}$, $-$31.2 km~s$^{-1}$, $-$28.6 km~s$^{-1}$, $-$26.7 km~s$^{-1}$, $-$23.6 km~s$^{-1}$, $-$19.7 km~s$^{-1}$, $-$19.1 km~s$^{-1}$ and $-$18.5 km~s$^{-1}$. These features conjure us an image of a bipolar mass-ejection from the openings of a torus.

Figure~6 shows the position-velocity (PV) diagrams of the CO~$J=2$--1 line. The cuts used for the PV diagrams are taken in the direction of elongation of the CO structure (120$^{\circ}$; corresponding to the symmetric axis of the torus) and a direction perpendicular to it (30$^{\circ}$). If the morpho-kinematic properties of the envelope are spherically symmetric, the two PV diagrams should exhibit the same pattern, but we see a difference between the two PV diagrams, suggesting that the central structure is spherically asymmetric. If the CO emission source consists only of a torus and expanding sphere (AGB wind), the PV diagrams must show two parallel slopes in PA=$120^{\circ}$ and an elliptical ring in PA=$30^{\circ}$ as we modeled in \citet{nak09}. Even though we see the sign of such slopes and ring in Figure~6, those are vague and rather complicated. Thus, one may think that another component complicates the circumstellar dynamics of IRAS 22272+5435. We consider this possibility in following sections.

%%%%%%%%%%%%%%%%%
%% Morpho-Kinematic Modeling %%
%%%%%%%%%%%%%%%%%

\section{Morphokinematic Modeling with {\it Shape}}
In order to acquire a better understanding of the morphokinematic properties of IRAS 22272+5435, we have constructed two models using the {\it Shape} software package \citep{ste11}. {\it Shape} is a tool to create three-dimensional (3D) models of astronomical nebulae. It was originally developed by \citet{ste06} for the analysis of optical/infrared spectroscopic data of PNe, in which one can assume optically thin conditions. {\it Shape} has also been repeatedly applied to radio molecular line observations of post-AGB stars, PPNe and PNe, in which lines are also not very optically thick, under the assumption of a optically thin condition \citep[here, "not very optically thick" means $\tau < 1$; see, e.g.,][]{ima09,nak09,nak10}. {\it Shape} does not calculate the full radiative transfer equations, however, the latest version (version 4.0 and later) can handle velocity dependent absorption (i.e., we can handle any values of $\tau$ in each velocity channel). This new capability of {\it Shape} enables us to simulate the results of full line radiative transfer calculations under some assumptions (for example, we need to assume a realistic model geometry, density distribution, etc.), providing a potent method to model the morphology and kinematics of astronomical nebulae. The algorithms simulating radiative transfer in {\it Shape} are designed to be exceptionally fast and to minimize the allocation of computer memory usage. {\it Shape} uses a ray-casting algorithm instead of the standard methods of line radiative transfer calculations, such as Monte-Carlo and $\lambda$-iteration methods. (Details of the {\it Shape} algorithms can be found on the website\footnote{http://bufadora.astrosen.unam.mx/shape/}.) The main difference between the {\it Shape} algorithm and typical radiative transfer calculations is that the attributes (such as emission and absorption coefficients) at each point in space are inputs and not calculated by the code. Therefore, in those instances where we can reasonably assume the attributes from the observational data, {\it Shape} can be a powerful tool for investigating morpho-kinematic properties of the nebulae. In the case of post-AGB stars, such as IRAS 22272+5435, we can assume a relatively simple geometry and density distribution. Additionally, our purpose is to discuss morpho-kinematic properties rather than deriving physical parameters of molecular gas (gas mass, etc.). In this case {\it Shape} can be a very useful tool. In our previous analyses using {\it Shape} \citep{nak09,nak10}, we had to assume optically thin conditions due to the limitations of previous versions of {\it Shape}, while in this study, with the new capability of the latest version, we have constructed a more realistic morpho-kinematic model.

\subsection{Model 1: Expanding Torus and AGB Spherical Wind}

% *** Figure~7: polygon-mesh of the torus and sphere (combined with the interaction region model). Create after parameters are fixed.
% Figure 8: channel maps and  PV diagrams of Model 1

As mentioned in Section 4, the CO emission from IRAS 22272+5435 cannot originate from merely an expanding torus and AGB spherical wind; an additional component or components are required to explain the observation. However, since we do not know the nature of these components, we first tried to fit the observational data only with a torus and spherically expanding sphere (we call this Model 1). This process with Model 1 allows us to determine the extent, to which the observations can be fitted with only these two components. Then, in the following subsection, we will try to model the entire CO structure by introducing an additional component as Model 2.  

According to Figure~2, the outermost region of the nebula deviates from spherical symmetry. Presumably, the outermost asymmetric structure is formed by the interaction between an AGB wind and the interstellar medium and/or resolving-out of largely extended emission by interferometry, and we do not attempt to model this asymmetric structure. In addition, we assumed an axial symmetric geometry in an effort to reduce the number of model parameters (for both Models 1 and 2). In the present modeling,  asymmetric structure (asymmetry with respect to the equatorial plane of the torus) is produced by the velocity dependent absorption of the outer spherical shell (i.e., emission of the inner structure is absorbed by the outer shell). The velocity of the approaching side of the inner structure, more or less, similar to that of the approaching side of the outer sphere, and therefore the emission from the approaching side of the inner structure is selectively absorbed by the outer shell. In contrast, the emission from the receding side of the inner structure is not absorbed by the outer shell, because the velocity of the receding side of the inner structure is clearly different from that of the near side (i.e., approaching side) of the outer shell. 

A polygon-mesh image of Model 1 is presented in the upper panels of Figure~7. The modeled sphere has a fixed outer radius of 2.5$''$  with an expanding velocity of $V_{\rm sp}=0.6r$, where $r$ is the angular distance from the central star in arcsec and the unit of $V_{\rm sp}$ is km~s$^{-1}$. $V_{\rm sp}$ is calibrated so that $V_{\rm sp}(r=2.5'') = 10.5$~km~s$^{-1}$. Although this radial dependence implies that there is a constant acceleration of the AGB wind in the radial direction (or the ejection velocity has been going down with time), usually the molecular gas of the AGB wind component has reached a terminal velocity by the post-AGB phase. Instead, this linear velocity law is intended to represent the effect of 2-3 different velocity components within the spherical AGB wind (i.e., the linear law is a simplification of multiple velocity components). The assumption of multiple velocity components is required for several reasons. The maximum expanding velocity of the AGB wind is fixed by the line-width of the spectrum at 10.5~km~s$^{-1}$. Therefore, if we assume a constant velocity of 10.5~km~s$^{-1}$ throughout the expanding AGB sphere, the models cannot reproduce the asymmetry with respect to the systemic velocity that is seen in the PV diagrams and channel maps. As was discussed in \citet{nak09}, the asymmetry seen in the PV diagram originates because emission from the torus is absorbed by the spherical component. To reproduce this absorption effect, the sphere must include components with velocities lower than 10.5~km~s$^{-1}$. 

The torus was modeled with four parameters: inner radius, outer radius, thickness (height), and expansion velocity. The position angle and inclination of the torus is confined by the MIR observations of \citet{uet01}. The model parameters obtained are: inner radius of 0.4$''$, outer radius of 1.0$''$, thickness of 0.6$''$, inclination of the symmetric axis of 50$^{\circ}$, position angle of the symmetric axis of 120$^{\circ}$, and a constant radial expansion velocity of 7.5 km~s$^{-1}$. The model parameters were determined by (educated) trial and error until the reproduced maps closely matched the observation. The model channel maps and PV diagrams were finally convolved with the synthesized beam pattern matching that of the observations ($-1.1''\times1.0''$ with PA$=-80.0^{\circ}$) for comparison with the observational maps.

In order to take into account absorption as a function of LSR velocity, we used the {\it Shape} physics module. The emission and absorption coefficients are defined as Gaussian functions:

\begin{equation}
j_{\lambda} = e^{-(\lambda - b)^2/(2c^2)} \times se \times n
\end{equation}
\begin{equation}
k_{\lambda} = e^{-(\lambda - b)^2/(2c^2)} \times sa \times n
\end{equation}

where $b$ is the central wavelength of the line, $c$ is the width of the line, $n$ is the {\it Shape}'s number density, $se$ is a multiplicative factor for the emission coefficient and $sa$ is a multiplicative factor for the absorption (Note: the {\it Shape}'s number density $n$ is the {\it Shape}'s internal parameter controlling the emissivity, and in the present case it is not directly related to the number density of CO molecules, as we stated later in Section 5.2).  Here $n$, $se$, and $sa$ all adjust the magnitude of the emission/absorption.  However, we use $se$ and $sa$ to adjust the overall emission/absorption, and $n$ is constrained to lie between 0 and 1.  The use of $n$ is to introduce emissivity gradients within the nebula.  By adjusting the $sa$ parameter, we can therefore increase/decrease the absorption by the sphere. The width of the line was set low enough as not to interfere with the line broadening due to the Doppler effect.  We applied the absorption function only to the sphere and assumed that the torus is optically thin (Note: even if we assume self absorption within the torus, the result is almost the same with the case of assuming no self-absorption. This is firstly because in the present modeling we assumed velocity-dependent absorption, and secondary because within the torus, gas components do not hide each other due to velocity differences). We find that an $sa$ factor giving an optical depth of 0.675 provided the right amount of absorption to match the observations.

% Description about the model maps.
In Figures~8 and 9, we present the channel maps and PV diagrams of Model~1, respectively, and in Figures~10 and 11, we present the difference between the observation and Model~1. It is obvious that all CO features can not be fit by this model. However, we see that two intensity peaks of the rim-brightened torus are produced by Model~1 in channels ranging from $-$26.1~km~s$^{-1}$ to $-$24.8~km~s$^{-1}$. In the case of the PV diagram, however, Model~1 fails to reproduce the observation; this supports that it is likely there are additional components other than the torus and sphere.

\subsection{Model 2: Interaction Region Assuming an Invisible Jet}

In Figures 10 and 11, we see some dense black regions, which is the residual not reproduced by Model 1, along with the symmetric axis of a torus (for example, see channels $-31.2$~km~s$^{-1}$, $-30.5$~km~s$^{-1}$, $-23.6$~km~s$^{-1}$ and $-22.9$~km~s$^{-1}$). Even though these features lying on the symmetric axis (or, on the directions of the opening of the torus) are reminiscent of a bipolar jet, unfortunately there are no clear-cut evidences proving the existence of a bipolar jet (for example, clearly collimated bipolar structure and acceleration along with the symmetric axis, etc.). However, a bipolar jet or jet-related activity is still a favorable explanation, since the presence of a jet is common in the early PPN phase. Therefore, we extended our modeling under the assumption that the emission unaccounted for by Model 1 originates from a hydrodynamical interaction between an ``invisible'' bipolar jet and the ambient material (torus and AGB wind). This ``invisible jet'' scenario may exist if the temperature of the jet is relatively high. In fact, the intensity of the CO $J=1$--0 and 2--1 lines tends to decrease as soon as the temperature reaches $\sim$50~K due to the population of higher-$J$ levels \citep[see, e.g.,][]{buj97,buj08}, while according to mid-infrared imaging \citep{uet01} the dust temperature of the inner regions of circumstellar envelopes could be far greater than 50~K.

We modeled the interaction region using a peanut-shaped shell with the emissivity enhanced in the polar regions to mimic a bow-shock structure. We constructed this geometry using the 3D editor in {\it Shape}. We started with a spherical shell with an outer radius of 1.3$''$ and an inner radius of 1.04$''$. We then pinched the waist of this sphere such that the equatorial radius was reduced to 0.52$''$ and linearly increased with distance along the symmetry axis to 1.3$''$ at the poles.  The result is a smooth bipolar, peanut shaped nebula. The velocity of the interaction region, like the torus, is set to 7.5 km~s$^{-1}$.  The center of the peanut-shaped shell is a cavity with $n=0$.

In order to create the bow-shock structure of the interaction region, we applied an emissivity gradient in the $\phi$ direction of $n(\phi)=(2 \phi/\pi-1)^8$.  Therefore at the polar regions where $\phi=180^{\circ}$, we reach a maximum of $n=1$ which sharply drops as $\phi$ is decreased.  A background emissivity given by $n=0.2$ is added to the rest of the shell (most of the emissivity is in the poles, but we added a small background emissivity to the rest of the bipolar shell, simulating, perhaps, the gas flowing from the bow-shock).  The position and inclination angles of the interaction region (i.e., symmetric axis of the structure) is 120$^{\circ}$ and 50$^{\circ}$ (although these angles are independently determined, the values are consequently the same with those of the torus). 

We added the modeled interaction region to Model 1 and called the result Model~2. Therefore, except for the interaction region, the definition of the torus and sphere in Model~2 is the same as for Model~1. The polygon-mesh image of Model~2 is presented in the right panel of Figure~7. The absorption settings in the {\it Shape} physics module are also the same as Model 1; we took into account the absorption only of the sphere, and we assumed that the torus and interaction region are optically thin (Note: as well as Model 1, even if we assume self absorption within the torus and interaction region, the result is almost the same with the case of assuming no self-absorption, because within the torus and interaction region, gas components almost do not hide each other due to velocity differences). 

In Figures~12 and 13, we present the channel maps and PV diagrams of Model 2, respectively. We find that the channels maps of Model~2 is much closer to the observational maps than those of Model~1 and additionally Model~2 is able to reasonably explain the observational PV diagrams. Since we assumed an axial symmetric geometry, we cannot reproduce the asymmetric morphology with respect to the symmetric axis of the torus and interaction region. For example, some intensity peaks seen Figure~5 (see, e.g., $-33.7$~km~s$^{-1}$, $-29.9$~km~s$^{-1}$ and $-27.4$~km~s$^{-1}$) are not produced in Model~2. However, as we reasonably reproduced the PV diagram with Model~2, the velocity dependent absorption of the outer shell seems to be a predominant reason explaining the asymmetry in the north--west to south--east direction. To explain the asymmetry with respect to the symmetric axis, presumably we need to assume, for example, asymmetric expansions and/or asymmetric density and temperature distributions. The Model 2, of course, may not be a unique solution to explain the observation, but at least the present analysis strongly suggests that an additional component other than the torus and AGB spherical wind is required to explain the observation.

One may think that the relatively large intensity of the interaction region (compared to the torus) is not consistent with the expected density distribution, which should be higher in the equatorial region than in the polar region. However, we shall clarify that we observed in the CO $J=2$--1 line, which is sensitive to a particular low range of gas temperature. As we stated in Section 5.2, the intensity of the CO $J=2$--1 line tends to decrease as soon as the temperature reaches $\sim$50~K due to the population of higher-$J$ levels. Therefore, a large intensity in the CO $J=2$--1 line does not immediately mean a high density. As the size of the torus seems to be smaller than that of the interaction region (i.e., closer to the central star), the torus is expected to exhibit a higher temperature than the interaction region. In fact, the torus is very clearly detected in mid-infrared imaging, which seems to be sensitive to a gas with a relatively high-temperature (say, $>$200--300~K), while the interaction region is not detected in mid-infrared imaging. This fact suggests that the interaction region exhibits a lower temperature than the torus. Therefore, there is no inconsistency even if we assume that the invisible jet with a relatively high temperature (faint in the $J=2$--1 line) is collimated by a hydrodynamical interaction with the dense torus with a relatively high temperature (faint in the $J=2$--1 line) and we see an interaction region region (with a relatively low temperature), which is bright in the CO $J=2$--1 line, along with the symmetric axis.  In addition, one may think that the interaction region is likely to have a higher velocity than the ambient wind. Indeed, at the tip of the invisible jet, the velocity may exhibit a higher than that of ambient material. However, as the CO $J=2$--1 line traces a relatively low temperature region, which presumably locates at the outermost part of the interaction layer, it is not unnatural even if the velocity is not extremely high. As known as a long standing puzzle in the field, the origin of the invisible jet itself is not clear. Interestingly, however, \citet{uet01} suggested the existence of a small post-AGB wind (angular size$\sim0.07''$), which is partially resolved by their high-resolution mid-infrared imaging. This structure of the post-AGB wind is not detected in the present observation (due presumably to the high temperature of the wind and the limitation of the angular resolution). However, if a part of this post-AGB wind is collimated by the dense torus, it could be a source to create the invisible jet.

%%%%%%%%%%%%%%%%%
%% Discussion %%
%%%%%%%%%%%%%%%%%

\section{Discussion}

The present CO interferometric observation and {\it Shape} modeling revealed the detailed morpho-kinematic properties of IRAS 22272+5435 for the first time. Here, we compare the morpho-kinematic properties of IRAS 22272+5435 and  IRAS 07134+1005 to clarify similarities and differences between the two prototypical 21~$\mu$m sources.

\subsection{Dynamical Timescale of the Torus}

The angular size of the torus determined in our modeling can be translated into linear sizes if we assume a distance of 1.67~kpc: the linear sizes of the inner radius, outer radius and thickness of the torus are calculated to be $1.0\times10^{16}$~cm, $2.5\times10^{16}$~cm and $1.5\times10^{16}$~cm, respectively. Unfortunately, the thickness may include large uncertainty, because the existence of the interaction region makes the shape of the torus indistinct particularly at the locations distant from the equatorial plane, whereas the edge of the inner and outer radii of the torus seem to be relatively clearly determined. Using the expanding velocity of 7.5~km~s$^{-1}$, the dynamical time-scales of the inner and outer edge of the torus are calculated to be 420 years and 1100 years, respectively. If we assume that the torus is formed by the superwind (equatorial enhanced mass-loss appeared in late AGB), these timescales suggest that the central star went into the superwind phase about 1100 years ago, and that the duration of the superwind was about 680 years. Then, finally the central star left the AGB 420 years ago; this value is consistent with previous IR studies \citep{uet01} suggesting that the central star left the AGB about 380 years ago.

In the case of IRAS 07134+1005, \citet{nak09} revealed that the linear size of the torus is somewhat larger than that of IRAS 22272+5435: the inner radius, outer radius and thickness of the torus of IRAS 07134+1005 are $4.3\times10^{16}$~cm, $3.0\times10^{17}$~cm and $4.5\times10^{16}$~cm, respectively: Putting aside the thickness including uncertainty (in the case of IRAS 22272+5435), the sizes of the inner and outer radii of the torus are several times larger than those of IRAS 22272+5435. The difference of the observed lines---\citet{nak09} observed the CO $J=3$--2 line, while  the CO $J=2$--1 line was observed in the present research---makes this more definite. If the IRAS 22272+5435 is observed in the higher-$J$ line, the angular size could be smaller than the present result, because the higher-$J$ line seems to trace the inner region with a higher temperature. 

Since both IRAS 22272+5435 and IRAS 07134+1005 exhibit almost the same expanding velocity of tori (expanding velocities of the tori of IRAS 22272+5435 and IRAS 07134+1005 are 7.5~km~s$^{-1}$ and 8.0~km~s$^{-1}$, respectively), IRAS 22272+5435 exhibits a relatively shorter dynamical time-scale of the torus. In fact, the timescale of the inner edge of IRAS 07134+1005 is 1140--1710 years \citep{nak09}. The smaller size and shorter time-scale of the torus of IRAS 22272+5435 is consistent with the result of single-dish observations in the CO $J=2$--1 and $J=4$--3 lines \citep{hri05}: the one-dimensional radiative transfer modeling based on the single-dish observations \citep{hri05} suggests that the mass-loss rate of IRAS 22272+5435 has sharply increased in the last 1000 years. On the contrary, the mass-loss rate of IRAS 07134+1005 is moderately increased compared to IRAS 22272+5435. The different time-scales of the inner radii of IRAS 22272+5435 and IRAS 07134+1005 are consistent with this fact.

\subsection{Interaction Region and Invisible Jet}

The most notable result found in the present observation is that CO emission of IRAS 22272+5435 cannot be explained only with an expanding torus and spherical AGB wind. Although a bipolar jet does not seem to be directly detected in the present CO observation, as suggested in Section 5.2 an interaction region between an invisible jet and ambient materials may explain the observation. Here we briefly consider the consistency on this idea.

As we modeled in Section 5.2, the length and thickness of the interaction region is 2.6$''$ and 0.3$''$, respectively. Therefore the distance from the central star to the inner edge of the interaction region is 1.0$''$, which corresponds to $2.5\times10^{16}$~cm at 1.67~kpc. If we assume an expanding velocity of the invisible jet that is currently interacting with ambient materials, we can roughly estimate the dynamical timescale of the invisible jet. As a template of the invisible jet, here we assume kinematic parameters of molecular jets found in water fountains, which are young post-AGB stars with oxygen-rich chemistry. Except for the difference in chemical properties, the evolutionary status of water fountains are quite similar to 21~$\mu$m sources. The tiny molecular jets in water fountains have been mapped with VLBI techniques in maser lines \citep[see, e.g.,][]{ima02,yun11} and the projected expanding velocities are known in a dozen of water fountains \citep{ima07}. According to \citet{ima07}, the projected jet velocities of water fountains are distributed between 50~km~s$^{-1}$ and 200~km~s$^{-1}$. If we assume this projected velocity as the jet velocity of the invisible jet in IRAS 22272+5435, the dynamical timescale of the invisible jet is calculated to be roughly 40--160 years. As the dynamical timescales of water fountain jets are distributed from 15--100 years \citep{ima07}, the dynamical timescale of the invisible jet in IRAS 22272+5435 is not inconsistent with the case of water fountains. The estimated dynamical time-scales of the torus and invisible jet in IRAS 22272+5435 is also consistent with \citet{hug07}, which suggested that the time lag between the torus and jet formation is in the range of 130 -- 1610 years (median is 300 years). According to our dynamical analysis, the torus and jet were started forming from 1100 years ago and 40--160 years ago, respectively. Therefore the time lag is 940--1060 years; this is within the range of the time lag suggested by \citet{hug07}, although the value is somewhat larger than the median. Of course, we shall note that the above discussion is very rough, and a large uncertainty could be included. For example, the tip of the interaction region most likely does not exactly correspond to the tip of the jet itself. The jet itself is likely to be considerably faster than the bow-shock. To make more precise discussions, of course, we need to directly detect the jet itself in future.

%%%%%%%%%%%%%
%% Summary %%
%%%%%%%%%%%%%

\section{Summary}
In this paper, we have reported the results of a CARMA CO observation of IRAS 22272+5435 in the CO $J=2$--$1$ line. We also performed morphokinematic analyses with {\it Shape}. The main results of this research are summarized below:
\begin{enumerate}
\item The emission distribution of the  CO $J=2$--$1$ line is not consistent with MIR structure. Even though both MIR and CO images exhibit two intensity peaks, the lines connecting the two peaks in MIR and CO images are almost perpendicular. This result is clearly different from the case of IRAS 07134+1005 \citep{nak09}, in which the CO structure is well accorded with the MIR structure of a rim-brightened torus.
\item A model based on the rim-brightened torus suggested from MIR observations explains a part of observational CO features, but a large deviation from a torus (plus sphere) model is found in the observational map. Although MIR images have been reasonably explained only with a torus and outer sphere, the present result suggests that, in addition to a torus and outer sphere, another component  may be included in the molecular envelope. 
\item The assumption of the interaction region between an invisible jet and ambient materials seems to reasonably explain the observation. In addition, the estimated dynamical time-scales of the torus and invisible jet are consistent with previous statistical studies on the formation of tori and jets in evolved stars.
\end{enumerate}

The invisible jet, of course, should be confirmed in follow-up observations. A key to directly detect the invisible jet would be to observe in the CO high-$J$ lines or vibrationally excited lines with a high-excitation temperature, because the small jet close to the central star could have a relatively high temperature. In fact, in the case of water fountains, the CO profiles of high-$J$ lines show Gaussian-like tails suggesting the existence of a high-velocity component \citep{he08,ima09}.

%%%%%%%%%%%%%%%%%%%%%
%% Acknowledgments %%
%%%%%%%%%%%%%%%%%%%%%

\acknowledgments
This work is supported by a grant awarded to JN and SN from the Research Grants Council of Hong Kong (project code: HKU 704710P; HKU 703110P) and the Small Project Funding of the University of Hong Kong (project code: 201007176004). Support for CARMA construction was derived from the Gordon and Betty Moore Foundation, the Kenneth T. and Eileen L. Norris Foundation, the James S. McDonnell Foundation, the Associates of the California Institute of Technology, the University of Chicago, the states of California, Illinois, and Maryland, and the National Science Foundation. Ongoing CARMA development and operations are supported by the National Science Foundation under a cooperative agreement, and by the CARMA partner universities. The authors thank Hiroshi Imai for stimulating discussions and valuable comments.

%%%%%%%%%%%%%%%%
%% references %%
%%%%%%%%%%%%%%%%

%%%%%%%%%%%%%%%%%%%%%%%
%%%%  Table %%%%%%
%%%%%%%%%%%%%%%%%%%%%

%\input{tab1.tex}

%\clearpage

%\input{tab2.tex}

%\clearpage

%%%%%%%%%%%%%
%% figure %%%
%%%%%%%%%%%%%

\begin{figure}
\epsscale{.60}
\plotone{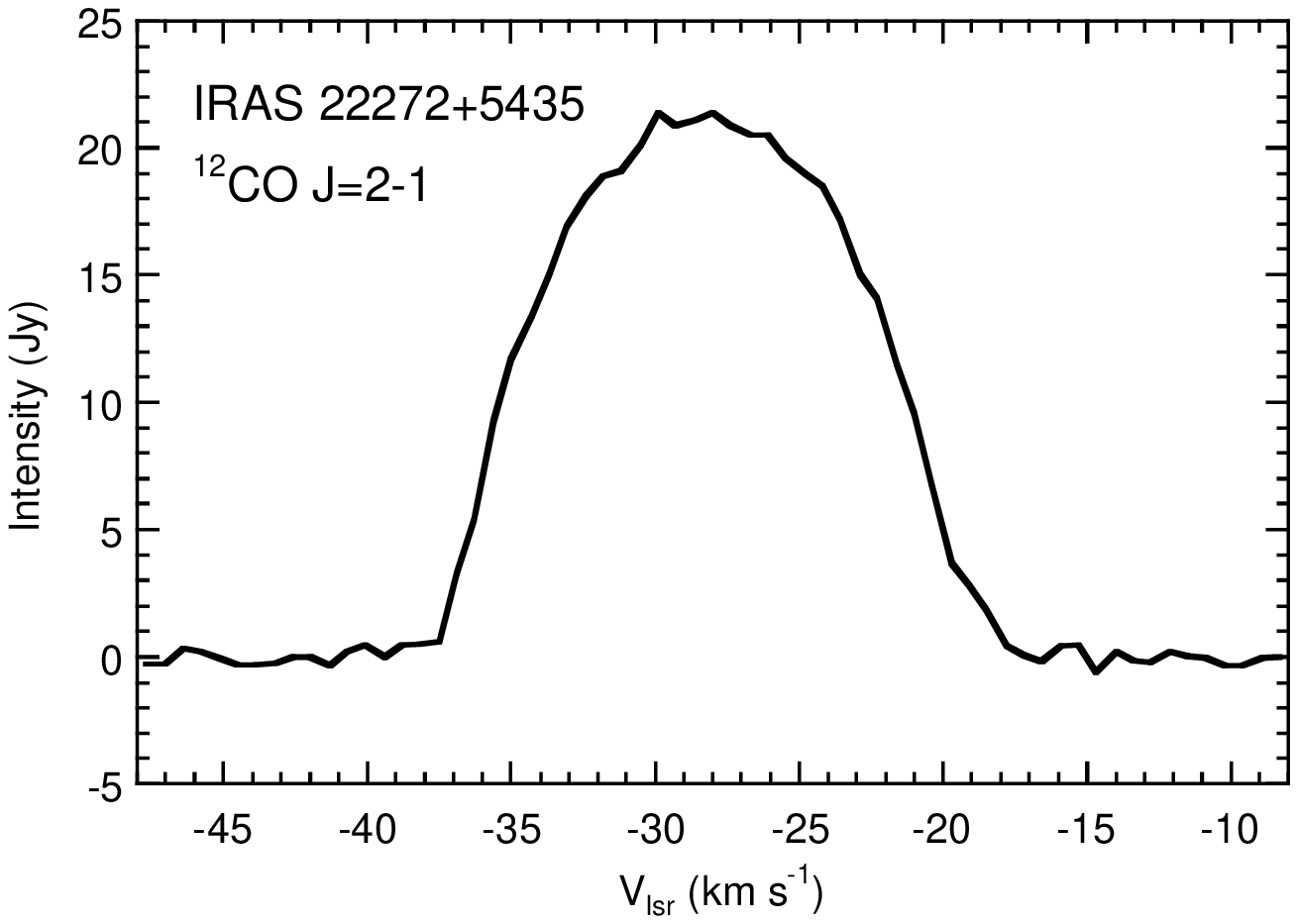}
\figcaption{$^{12}$CO $J=2$$-$$1$ total flux line profile. \label{fig1}}
\end{figure}
\clearpage

\begin{figure}
\epsscale{.80}
\plotone{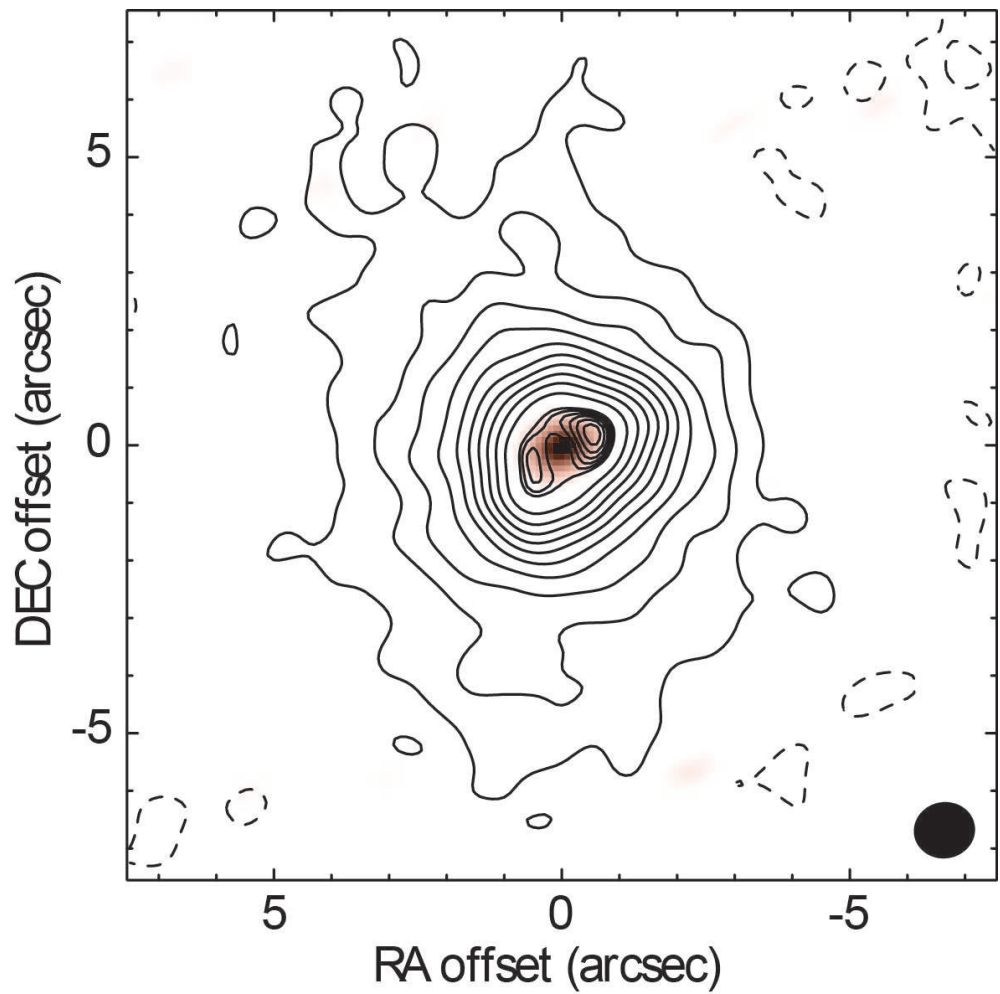}
\figcaption{Total flux intensity map in the $^{12}$CO $J=2$$-$$1$ line superimposed on the 1~mm radio continuum image. The contours start from the 3~$\sigma$ level, and the levels are spaced every 4~$\sigma$ until  the 43~$\sigma$ level, and above the 43~$\sigma$ level the levels are spaced every 0.5~$\sigma$. The highest contour corresponds to the 45.5~$\sigma$ level. The 1~$\sigma$ level corresponds to $1.77\times10^{-2}$ Jy~beam$^{-1}$. The dashed contour correspond to $-$3~$\sigma$. The FWHM beam size is located in the bottom right corner. The origin of the coordinate corresponds to the phase center. \label{fig2}}
\end{figure}
\clearpage

\begin{figure}
\epsscale{.90}
\plotone{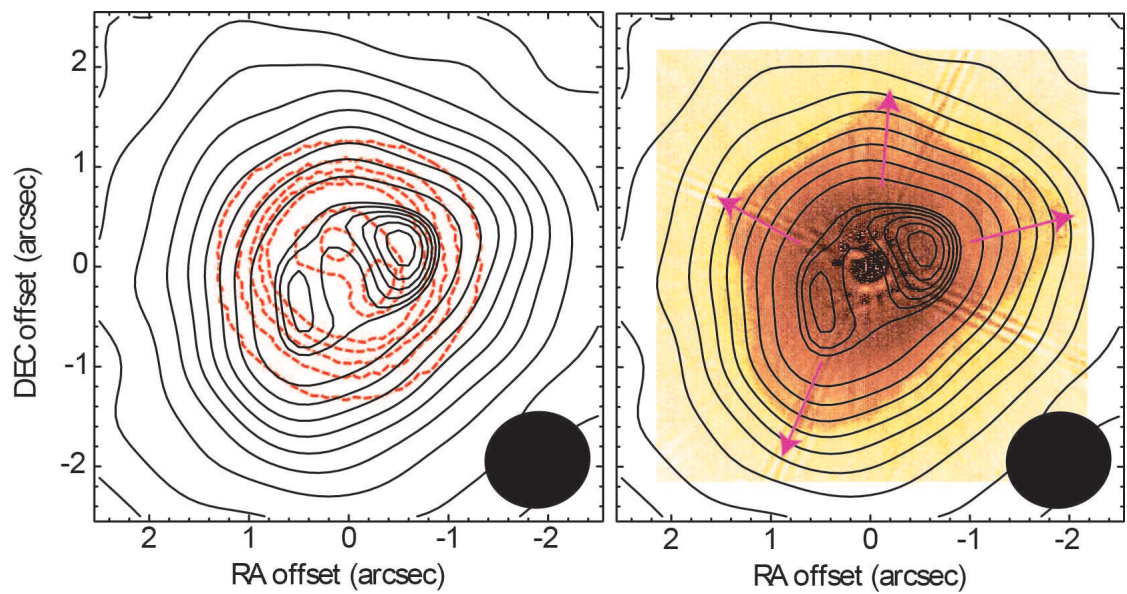}
\figcaption{{\it Left}: Total flux intensity map in the $^{12}$CO $J=2$$-$$1$ line superimposed on the mid-infrared 12.5~$\mu$m image taken from (Ueta~et~at.~2001). The contour levels are the same as Figure~2. The FWHM beam size is located in the bottom right corner. {\it Right}: Total flux intensity map in the $^{12}$CO $J=2$$-$$1$ line superimposed on the $HST$ $I$--band image taken from (Ueta~et~at.~2000). The contour levels are the same as Figure~2. The FWHM beam size is located in the bottom right corner. The pink arrows indicate the directions of the elliptical protrusions suggested by \citet{uet01}. \label{fig3}}
\end{figure}
\clearpage

\begin{figure}
\epsscale{.90}
\plotone{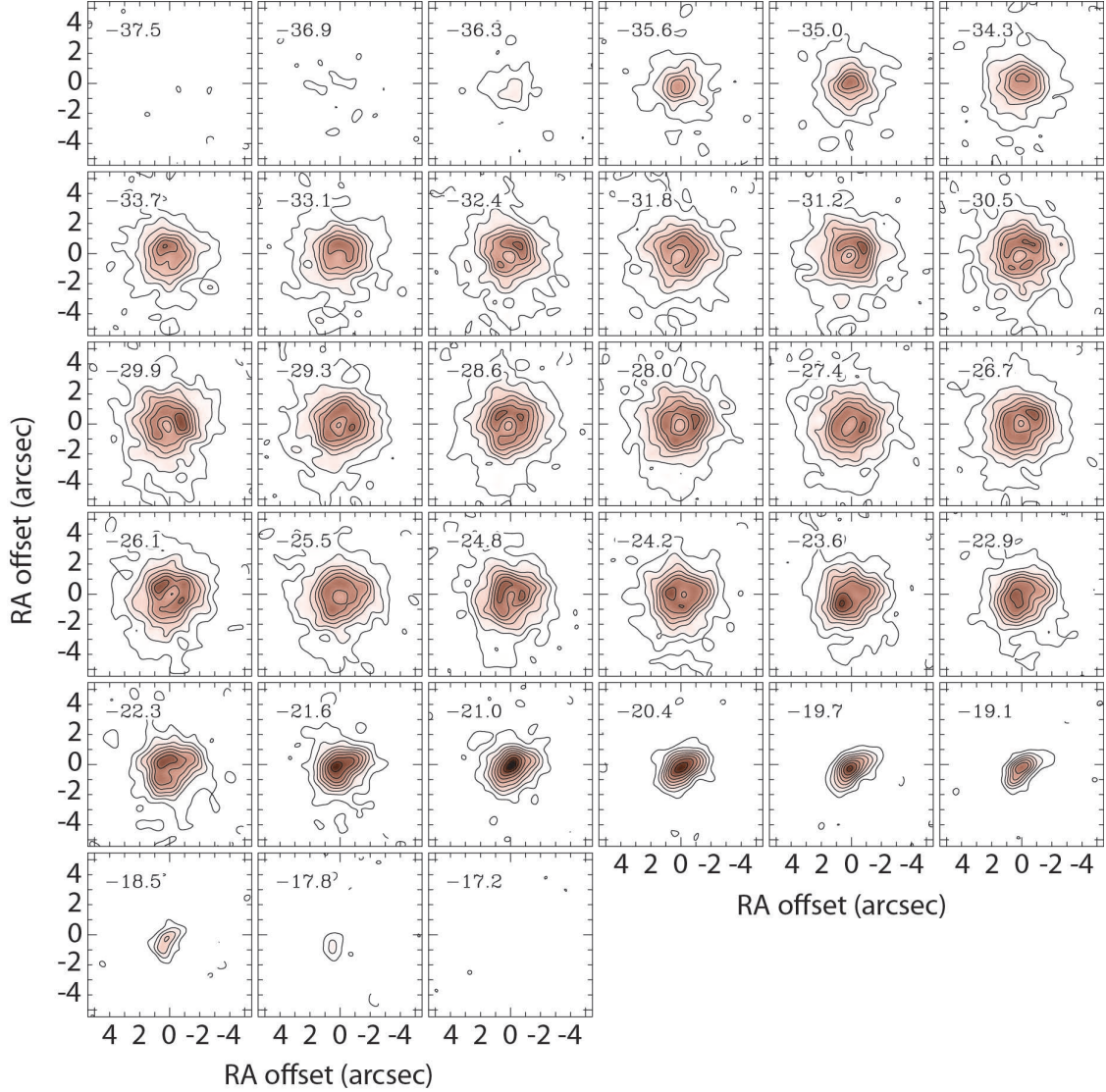}
\figcaption{Channel maps of the $^{12}$CO $J=2$$-$$1$ line. The velocity width of each channel is 0.635 km s$^{-1}$ and the central velocity in km s$^{-1}$ is located in the top left corner of each channel map. The contours start from the 3~$\sigma$ level, and the levels are spaced every 3~$\sigma$. The 1~$\sigma$ level corresponds to $5.62\times10^{-2}$ Jy~beam$^{-1}$. The dashed contour correspond to $-$3~$\sigma$ (almost no $-$3~$\sigma$ contours are seen in the map). The FWHM beam size is located in the bottom right corner. The origin of the coordinate corresponds to the phase center. \label{fig3}}
\end{figure}
\clearpage

\begin{figure}
\epsscale{.90}
\plotone{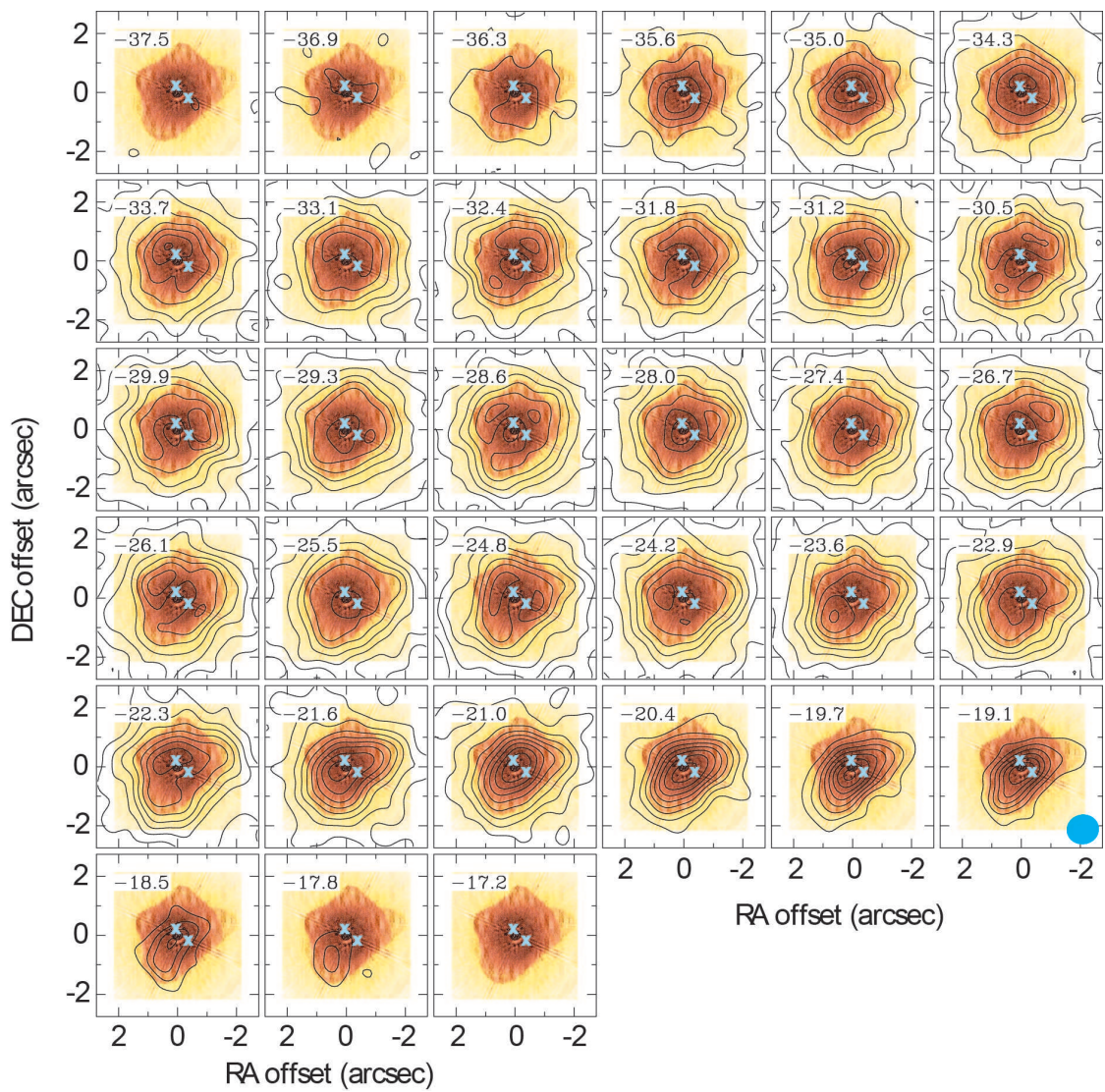}
\figcaption{Enlarged channel maps of the $^{12}$CO $J=2$$-$$1$ line superimposed on the $HST$ $I$--band image taken from (Ueta~et~at.~2000). The blue crosses represent the emission peaks of a rim-brightened torus suggested by \citet{uet01}. The other notations of the diagram is the same as Figure~4. \label{fig5}}
\end{figure}
\clearpage

\begin{figure}
\epsscale{.80}
\plotone{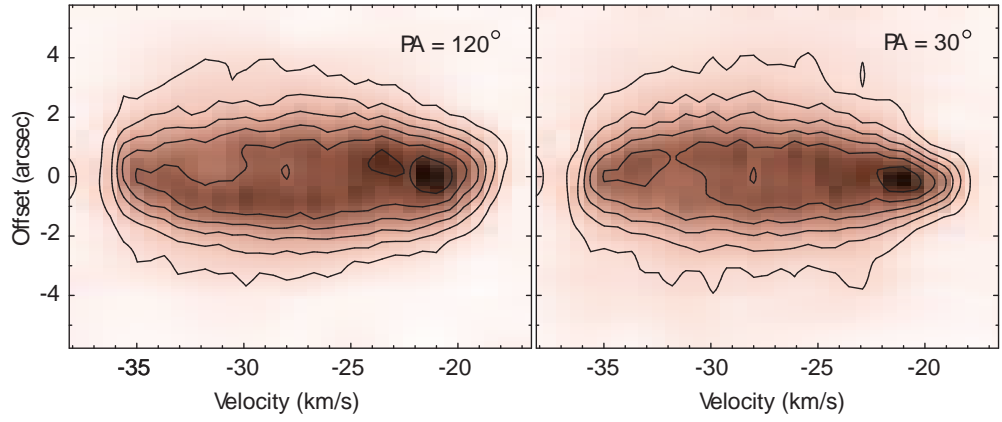}
\figcaption{Position--velocity diagrams of the CO $J=2$--1 line. The contour levels are 15, 30, 45, 60, 75 and 90\% of the intensity peak, and the peak intensity is 1.26 Jy~beam$^{-1}$. The position angles of the cuts are given in the upper-right corners. \label{fig6}}
\end{figure}
\clearpage

\begin{figure}
\epsscale{1.0}
\plotone{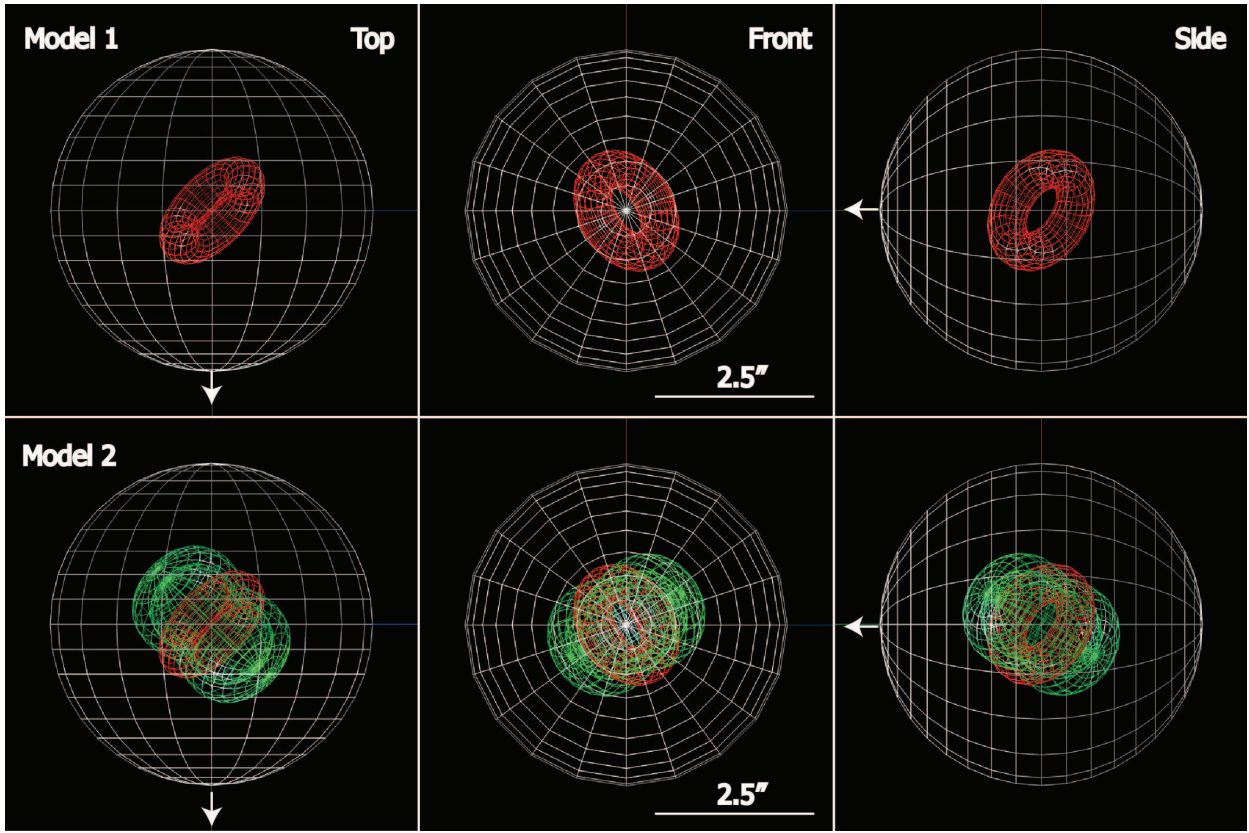}
\figcaption{Polygon-mesh images of the Shape models. Model 1 consists of  a torus and sphere, and Model 2 consists of a torus, sphere and axially symmetric interaction region (see, text). The red, green and white meshes represent a torus, interaction region and sphere, respectively. The central panels (Front) shows the line-of-sight views, the right panels (Side) show views from the east side, and the left panels (Top) show views from the north. The angular scales are given in the lower-right corners in the central panels. The white arrows represent the direction to the observer. \label{fig7}}
\end{figure}
\clearpage

\begin{figure}
\epsscale{1.0}
\plotone{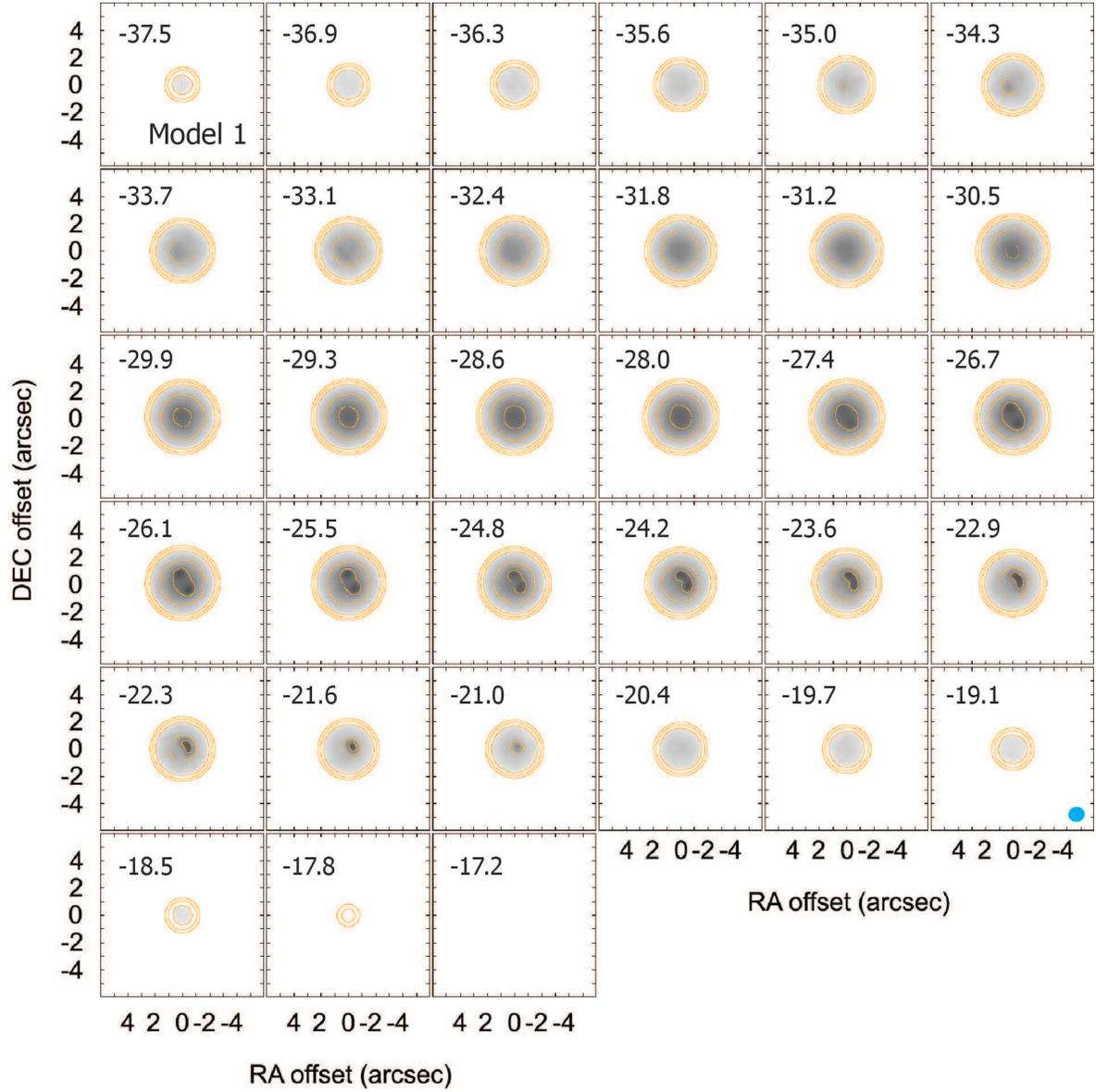}
\figcaption{Channel maps of Model~1 (consisting of an expanding torus and sphere; see text). The beam pattern used for convolution is located in the bottom right corner. \label{fig8}}
\end{figure}
\clearpage

\begin{figure}
\epsscale{.90}
\plotone{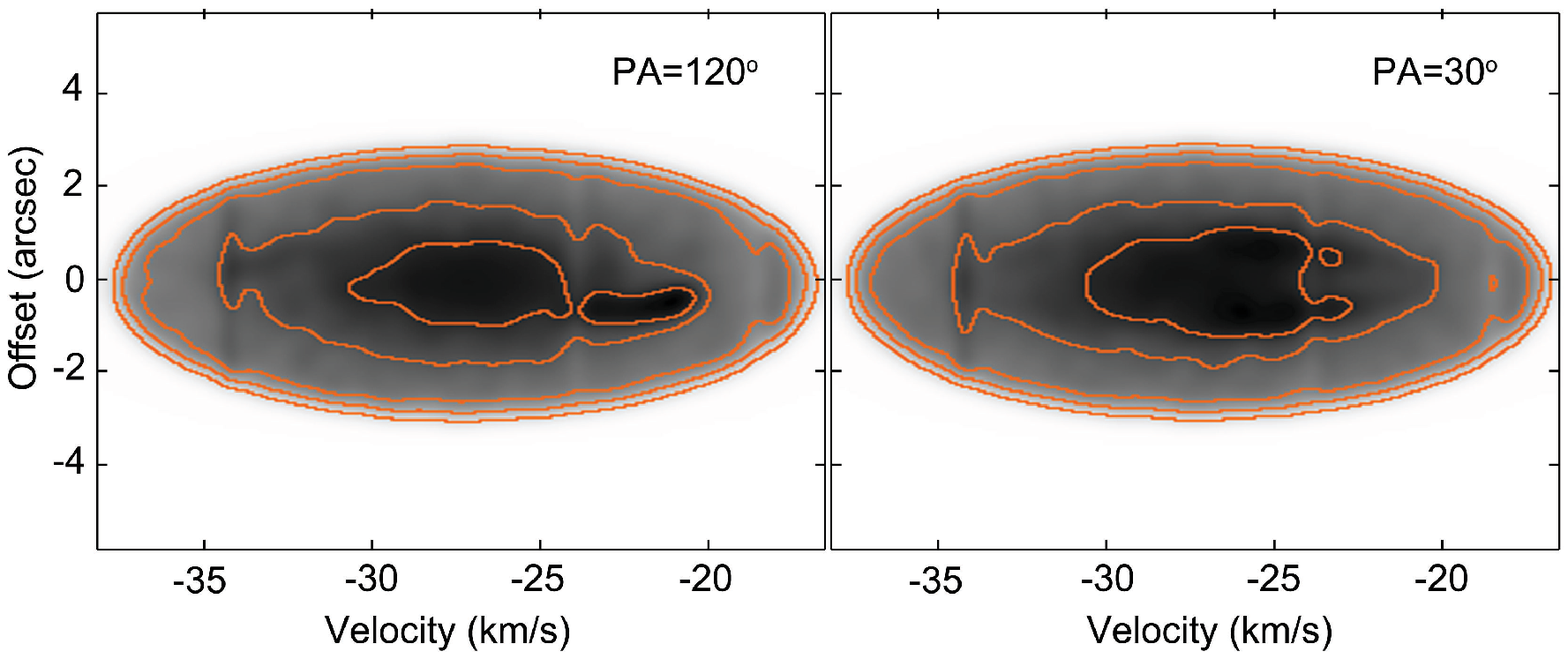}
\figcaption{Position-velocity diagrams of Model~1. The intensity distribution is convolved with the observational beam. \label{fig9}}
\end{figure}
\clearpage

\begin{figure}
\epsscale{.90}
\plotone{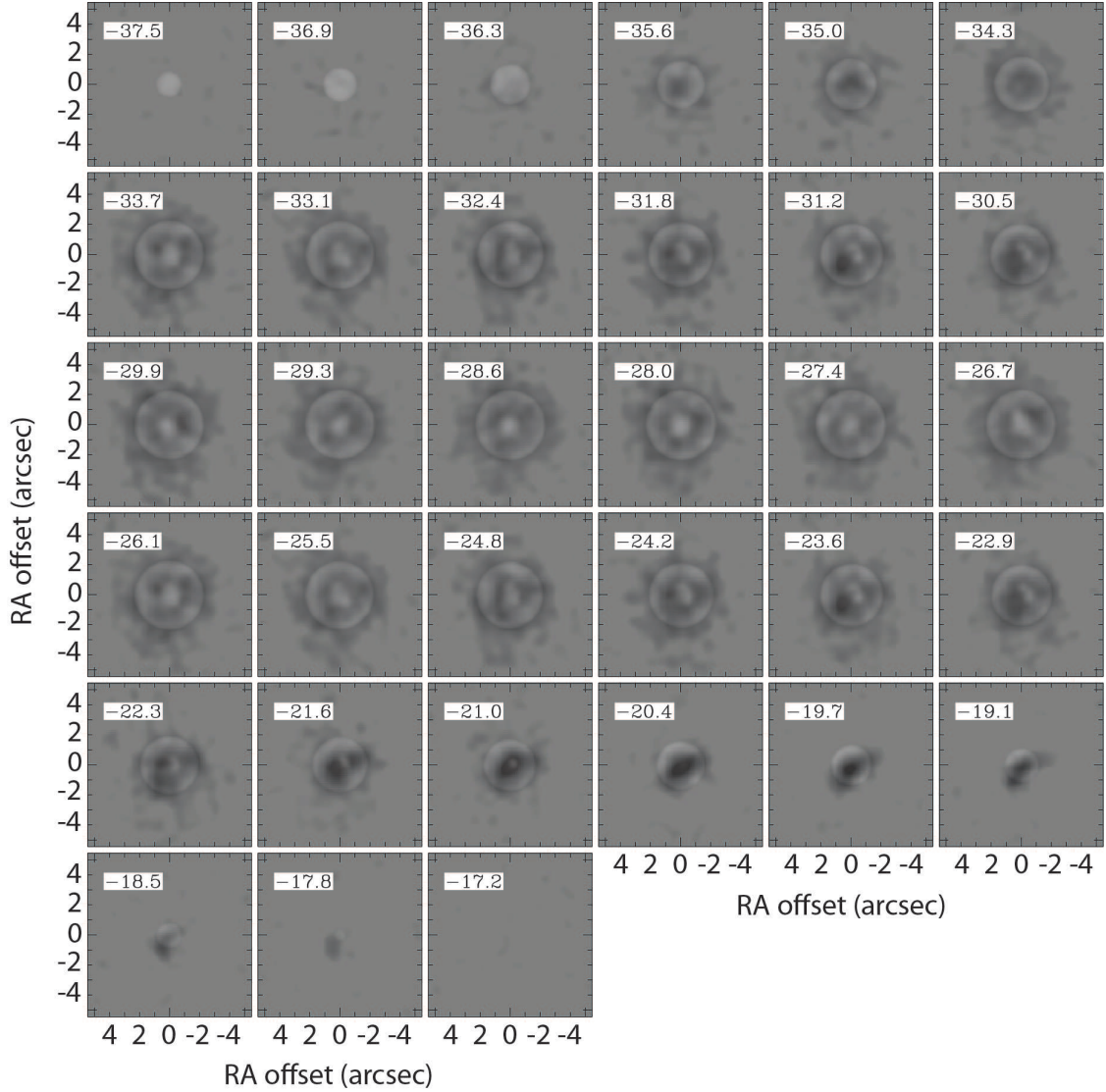}
\figcaption{Difference channel maps between the observation and Model~1. The model map is subtracted from the observational map after convolving with the observational beam. The white means the model is too bright, and the black means the observations are too bright (therefore, if it was a perfect match the entire image would be grey). \label{fig10}}
\end{figure}
\clearpage

\begin{figure}
\epsscale{.90}
\plotone{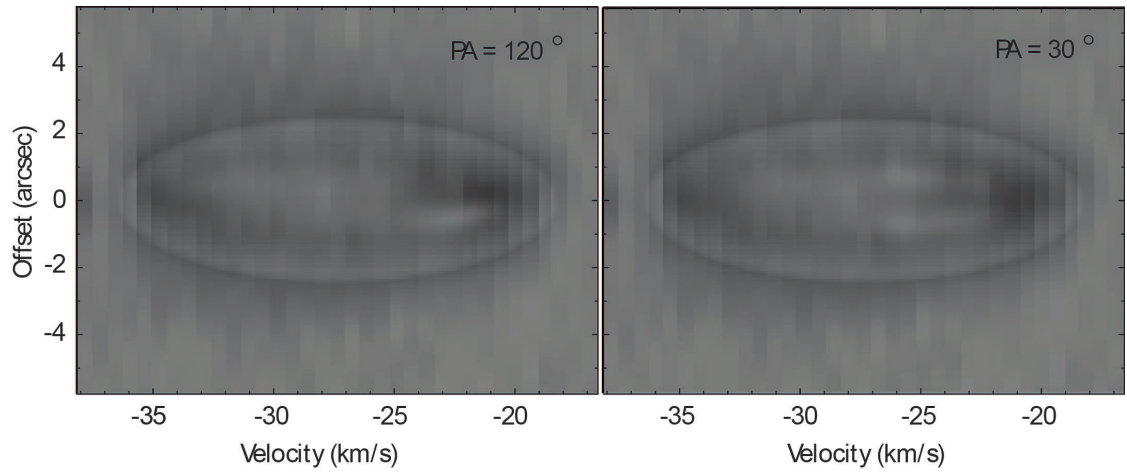}
\figcaption{Defference position-velocity diagrams between the observation and Model~1. The meaning of the gray scale is the same as Figure~10. \label{fig11}}
\end{figure}
\clearpage

\begin{figure}
\epsscale{1.0}
\plotone{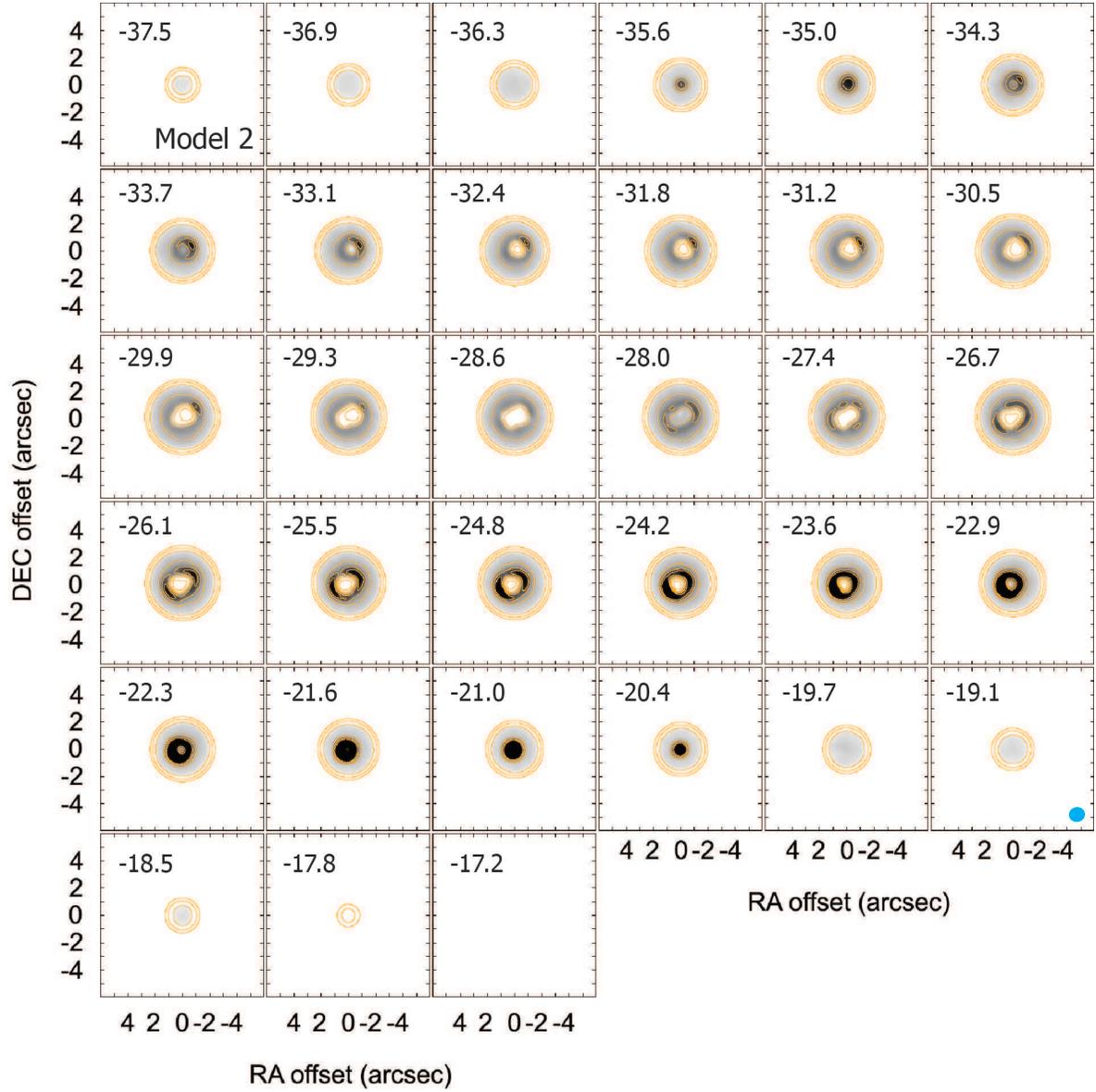}
\figcaption{Channel maps of Model~2 (consisting of an expanding torus, axisymmetric interaction region and sphere; see text). The beam pattern used for convolution is located in the bottom right corner. \label{fig12}}
\end{figure}
\clearpage

\begin{figure}
\epsscale{.90}
\plotone{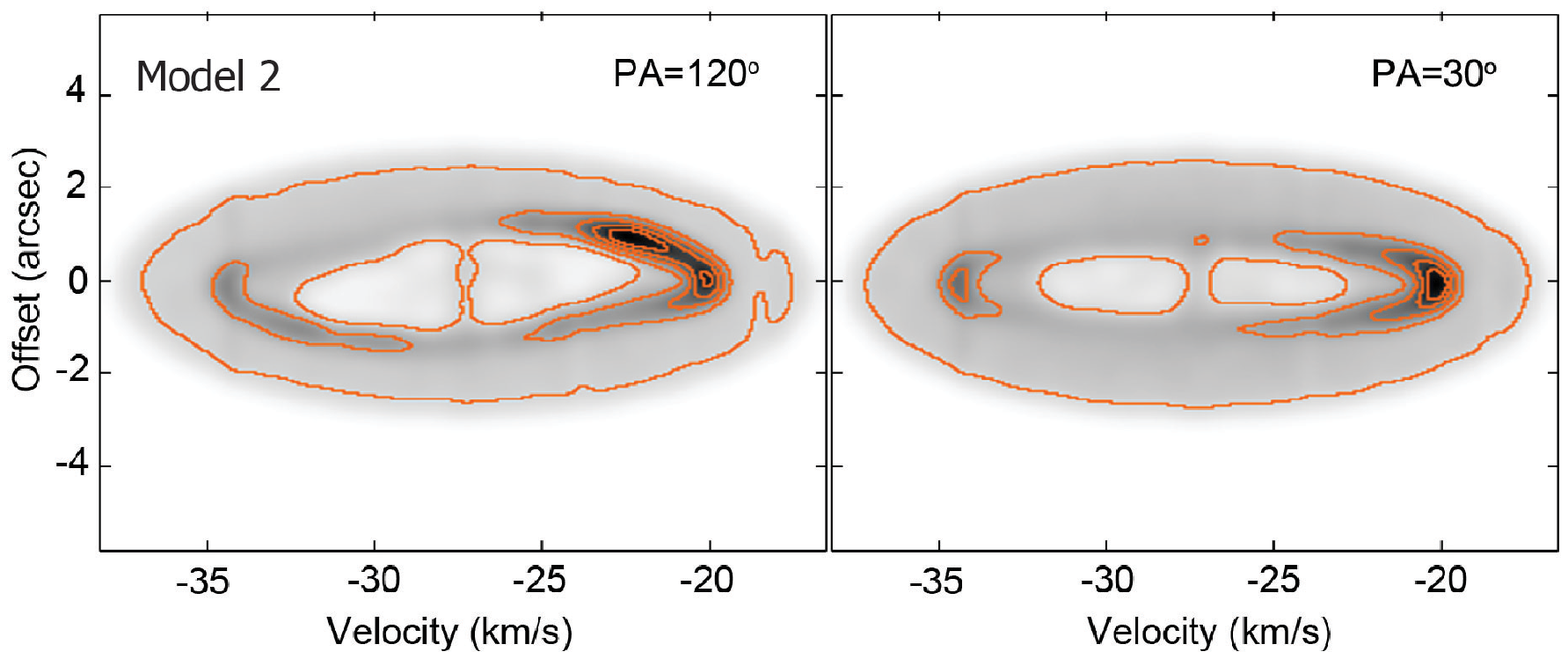}
\figcaption{Position-velocity diagrams of Model~2. The intensity distribution is convolved with the observational beam. \label{fig13}}
\end{figure}
\clearpage

%%%%%%%%%%%%%%%%%%%%%%%%%%%%%%
\end{document}